\documentclass{JHEP3}
\usepackage{graphicx}
\usepackage{latexsym}
\newcommand{\nn}{\nonumber\\}\newcommand{\p}[1]{(\ref{#1})}

\def\a{\alpha}\def\b{\beta}\def\g{\gamma}\def\d{\delta}\def\e{\epsilon}

\def\l{\lambda}

\def\PRL #1 #2 #3{{\em Phys. Rev. Lett. \/} {\bf#1} (#2) #3}
\def\NPB #1 #2 #3{{\em Nucl. Phys. \/} {\bf B#1} (#2) #3}
\def\NPBFS #1 #2 #3 #4{{\em Nucl. Phys. \/} {\bf B#2} [FS#1] (#3) #4}
\def\CMP #1 #2 #3{{\em Commun. Math. Phys. \/} {\bf #1} (#2) #3}
\def\PRD #1 #2 #3{{\em Phys. Rev. \/} {\bf D#1} (#2) #3}
\def\PLA #1 #2 #3{{\em Phys. Lett. \/} {\bf #1A} (#2) #3}
\def\PLB #1 #2 #3{{\em Phys. Lett. \/} {\bf B#1} (#2) #3}
\def\JMP #1 #2 #3{{\em J. Math. Phys. \/} {\bf #1} (#2) #3}
\def\PTP #1 #2 #3{{\em Prog. Theor. Phys. \/} {\bf #1} (#2) #3}
\def\SPTP #1 #2 #3{{\em Suppl. Prog. Theor. Phys. \/} {\bf #1} (#2) #3}
\def\AoP #1 #2 #3{{\em Ann. of Phys. \/} {\bf #1} (#2) #3}
\def\PNAS #1 #2 #3{{\em Proc. Natl. Acad. Sci. USA} {\bf #1} (#2) #3}
\def\RMP #1 #2 #3{{\em Rev. Mod. Phys. \/} {\bf #1} (#2) #3}
\def\PR #1 #2 #3{{\em Phys. Reports \/} {\bf #1} (#2) #3}
\def\AoM #1 #2 #3{{\em Ann. of Math. \/} {\bf #1} (#2) #3}
\def\UMN #1 #2 #3{{\em Usp. Mat. Nauk \/} {\bf #1} (#2) #3}
\def\FAP #1 #2 #3{{\em Funkt. Anal. Prilozheniya \/} {\bf #1} (#2) #3}
\def\FAaIA #1 #2 #3{{\em Functional Analysis and Its Application \/} {\bf
#1} (#2) #3}
\def\BAMS #1 #2 #3{{\em Bull. Am. Math. Soc. \/} {\bf #1} (#2)
#3} \def\TAMS #1 #2 #3{{\em Trans. Am. Math. Soc. \/} {\bf #1}
(#2) #3}
\def\InvM #1 #2 #3{{\em Invent. Math. \/} {\bf #1} (#2) #3}
\def\LMP #1 #2 #3{{\em Letters in Math. Phys. \/} {\bf #1} (#2) #3}
\def\IJMPA #1 #2 #3{{\em Int. J. Mod. Phys. \/} {\bf A#1} (#2) #3}
\def\AdM #1 #2 #3{{\em Advances in Math. \/} {\bf #1} (#2) #3}
\def\RMaP #1 #2 #3{{\em Reports on Math. Phys. \/} {\bf #1} (#2) #3}
\def\IJM #1 #2 #3{{\em Ill. J. Math. \/} {\bf #1} (#2) #3}
\def\APP #1 #2 #3{{\em Acta Phys. Polon. \/} {\bf #1} (#2) #3}
\def\TMP #1 #2 #3{{\em Theor. Mat. Phys. \/} {\bf #1} (#2) #3}
\def\JPA #1 #2 #3{{\em J. Physics \/} {\bf A#1} (#2) #3}
\def\JSM #1 #2 #3{{\em J. Soviet Math. \/} {\bf #1} (#2) #3}
\def\MPLA #1 #2 #3{{\em Mod. Phys. Lett. \/} {\bf A #1} (#2) #3}
\def\JETP #1 #2 #3{{\em Sov. Phys. JETP \/} {\bf #1} (#2) #3}
\def\JETPL #1 #2 #3{{\em  Sov. Phys. JETP Lett. \/} {\bf #1} (#2) #3}
\def\PHSA #1 #2 #3{{\em Physica} {\bf A#1} (#2) #3}
\def\CQG #1 #2 #3{{\em Class. Quantum Grav. \/} {\bf #1} (#2) #3}
\def\SJNP #1 #2 #3{{\em Sov. J. Nucl. Phys. (Yadern.Fiz.) \/} {\bf #1} (#2) #3}

\def\be{\begin{equation}}
\def\ee{\end{equation}}
\def\ba{\begin{array}} \def\ea{\end{array}}
\def\bea{\begin{eqnarray}}
\def\eea{\end{eqnarray}}
\def\nn{\nonumber}

\title {Higher Spins from Tensorial Charges and $OSp(N|2n)$
Symmetry}

\author{Mikhail Plyushchay\\
 Departamento de F\'{\i}sica,
Universidad de Santiago de Chile,
Casilla 307, Santiago 2, Chile \\
Institute for High Energy Physics,
Protvino, Russia \\ E-mail: \email{mplyushc@lauca.usach.cl}}

\author{Dmitri Sorokin \\Institute for Theoretical Physics, NSC KIPT,
61108 Kharkov, Ukraine \\
Dipartimento di Fisica, Universit\`a degli Studi di Padova,
and Istituto Nazionale di Fisica Nucleare, Sezione di Padova
Via F. Marzolo 8, 35131 Padova, Italy \\ E-mail:
\email{dmitri.sorokin@pd.infn.it}}
\author{Mirian Tsulaia \\
Dipartimento di Fisica, Universit\`a degli Studi di Padova,
and Istituto Nazionale di Fisica Nucleare, Sezione di Padova
Via F. Marzolo 8, 35131 Padova, Italy \\
Bogoliubov Laboratory of Theoretical Physics, JINR,
 141980 Dubna, Russia \\
Institute of  Physics, GAS, 380077 Tbilisi, Georgia \\
E-mail: \email{mirian.tsulaia@pd.infn.it}}

\abstract{It is shown that the quantization of a superparticle
propagating in an $N=1$, $D=4$ superspace extended with tensorial
coordinates results in an infinite tower of massless spin states
satisfying the Vasiliev unfolded equations for free higher spin
fields in flat and $AdS_4$ $N=1$ superspace. The tensorial
extension of the $AdS_4$ superspace is proved to be a supergroup
manifold $OSp\,(1|4)$. The model is manifestly invariant under
$OSp\,(N|8)$ (N=1,2) superconformal symmetry. As a byproduct, we
find that the Cartan forms of arbitrary $Sp(2n)$ and $OSp(1|2n)$
groups are $GL\,(2n)$ flat, i.e. they are equivalent to flat
Cartan forms up to an exactly determined $GL\,(2n)$ rotation. This
property is crucial for carrying out the quantization of the
particle model on $OSp\,(1|4)$ and getting the higher spin field
dynamics in super $AdS_4$, which can be performed in a way
analogous to the flat case.}


\preprint{USACH-FM-03-01, DFPD 03/TH/02}

\begin{document}

\section{Introduction}

It is well known that the higher spin fields naturally arise in
the quantization of classical superstrings and describe an
infinite tower of quantum string states with masses linearly
depending on spin and on inverse string tension. In the limit
where the string tension tends to zero all (higher spin) string
states become massless and may be regarded as gauge fields of an
infinite, so called, higher spin symmetry. There is a conjecture
that string theory is a spontaneously broken phase of a gauge
theory of higher spin fields. If this conjecture is realized, it
can be useful for better understanding of string/M theory and of
the AdS/CFT correspondence (see e.g. \cite{V,BO,DNW}). This is one
of the motivations of the development of the theory of interacting
higher spin fields.

Understanding the interactions of higher spin fields is a main
problem of the construction of the higher spin field theory, the
gauge invariance playing a crucial role in cancelling the
contribution of the states of negative norm. The interaction
problem already reveals itself when one tries to couple higher
spin fields to gravity. As it has been realized in \cite{FV} a
space--time with a constant curvature, in particular an Anti de
Sitter space, rather than a flat space, is an appropriate vacuum
background on which the construction of a consistent interacting
theory of higher spin fields should be based.

There are at least two alternative approaches to the description
of massless higher spin fields both in flat and in AdS backgrounds
(see e.g. \cite{segal} for a recent review and for the
construction of conformal higher spin theory). One of the
approaches is the formulation in terms of totally symmetric tensor
fields (see \cite{F1} for the case of $AdS_4$, \cite{Me,JB1} for the
case of arbitrary dimensional AdS spaces and \cite{sagnotti} for
the latest developments). Another approach is a geometrical
formulation \cite{FV}, \cite{V}, \cite{vas} -- \cite{SeS} in which
an important role in the construction of interacting higher spin
field theory is played by the algebra of higher spin symmetries
$hs(2n)$. Since the number of higher spin fields is infinite the
algebra $hs(2n)$ is infinite dimensional and contains the algebra of
isometries of the AdS space as a finite dimensional subalgebra. The
realization of the higher spin algebra on the higher spin fields
is described with the help of auxiliary spinor variables which are
very similar to twistors.

This fact may have quite interesting implications. One may ask if
there exists a (super)particle model whose quantization would
produce an infinite set of the higher spin fields and whether it
may be used to understand how the higher spin interactions can be
introduced. Since together with an AdS background, the consistent
interactions of the higher spin fields require that the whole
infinite tower of them be involved, the quantum states of such a
superparticle should form the infinite higher spin set, and not a
finite one, as conventional superparticle and spinning particle
models have. In other words, the theory should not have the
constraint which fixes the value of the second Casimir operator
(helicity). Twistors turn out to be rather useful to construct
such a model. In \cite{bl} there has been proposed a model of a
massless superparticle in an $N=1$, $D=4$ superspace enlarged with
tensorial coordinates and twistor--like spinor variables. The
conjugate momenta of these tensorial coordinates appear as
tensorial charges in a superalgebra which generates the
supersymmetry of the system. It is worth noting that this model
has been the first example of BPS states which preserve more than
one half supersymmetry. (A generalization of this model to a
massive superparticle has been discussed in \cite{fedor}. See also
\cite{preon,rubik} for further developments and applications to
the case of branes). The quantum spectrum of the massless
tensorial superparticle has been shown \cite{exotic} to consist of
an infinite tower of massless higher (integer and half integer)
spin states, while the twistorial variables become noncommutative,
and it has been assumed that this theory corresponds to the higher
spin field theory developed by Vasiliev \cite{vas,misha,misha1}.

The aim of the present paper is to study in more detail this
relationship and, in particular, to analyze the $OSp(1|8)$ and
$OSp\,(2|8)$ superconformal invariance of the model and to derive
the so called `unfolded' equations of motion of higher spin fields
\cite{vas,misha,misha1} which appear as a result of the
quantization of the superparticle.

We start with the superparticle model in the flat $N=1$, $D=4$
superspace extended with tensorial coordinates and determine its
$OSp(N|8)$ $(N=1,2)$ structure. We then pass to a superparticle in
an $N=1$ supersymmetric $AdS_4$ background and show that its
tensorial extension is a supergroup manifold $OSp(1|4)$. The
classical dynamics of a superparticle on super $AdS_4$ and on
$OSp(1|4)$ has already been considered in \cite{preit}.

At this point we should note that since the relationship of the
quantized superparticle models of \cite{exotic,preit} to the
unfolded higher spin field equations was not clear, in
\cite{misha} the actions for these superparticles were enlarged
with new variables along with kinetic terms for twistorial
coordinates. Such a modification changed the structure of first
and second class constraints but at the same time kept the number
of physical degrees of freedom of the models intact. This
generalization is similar to and is a Lagrangian form of the
conversion procedure used in \cite{exotic} to quantize the
superparticle by converting second class constraints into first
class constraints with the help of new auxiliary variables. In
\cite{misha} this allowed to carry out a BRST quantization of
these generalized superparticles which yielded unfolded equations
of motion of higher spin fields as functions of the coordinates of
the generalized tensorial superspace. The same equations, in a
momentum representation for spinorial variables, were obtained in
\cite{exotic}, but their meaning as unfolded equations was not
understood therein. A main result of \cite{misha} has been that
using arguments by Fronsdal \cite{fronsdal1}, who was actually the
first to realize the importance of simplectic group manifolds for
the description of higher spins, it was shown that in $D=4$ the
dynamics in the tensorial space is locally and globally equivalent
to the unfolded free higher spin field dynamics in flat Minkowski
and $AdS$ spaces.

 In the present paper we show that it is possible to avoid
 the introduction of new variables
and of the corresponding kinetic terms. The superparticle models
\cite{exotic,preit} are quantized in a (twistor) momentum
representation in contrast to the tensorial space coordinate
quantization performed in \cite{misha}. This allows us, upon
applying an appropriate Fourier transform, to get higher spin wave
functions propagating and satisfying the unfolded equations
directly in the physical Minkowski and $AdS$ subspaces of the
tensorial superspace.

We also demonstrate that both, the flat tensorial superspace and
$OSp(1|4)$ can be regarded as different coset superspaces of the
form ${{OSp(1|8)}\over{GL(4)\times\!\!\!\!\supset SK}}$, where
$GL(4)\times\!\!\!\!\!\!\supset SK$ is a semidirect product of the
general linear group $GL(4)$ and of a generalized super Poincare
group $SK$, which are subgroups of $OSp(1|8)$. As a consequence
the Cartan forms (or supervielbeins) of the two supermanifolds are
related to each other by a $GL(4)$ transformation. As a byproduct
we find that the Cartan one--forms of any  supergroup $OSp(1|2n)$
($n=1,\cdots,\infty$) are $GL(2n)$ flat, i.e. that they differ
from the Cartan forms of a corresponding flat tensorial superspace
by a definite explicitly found $GL(2n)$ rotation (see Section 6).
Such peculiar properties of $OSp(1|2n)$ and in particular of the
$OSp(1|4)$ supermanifold will allow us to quantize the
superparticle on $OSp(1|4)$ in a simple way by applying the
twistor technique used for the quantization of the superparticle
in flat tensorial superspace \cite{exotic} and adopted below to
the derivation of the unfolded higher spin field equations in
Minkowski and $AdS_4$ superspaces.

Our conclusion, which is in agreement with the philosophy of
\cite{fronsdal1,exotic,preit,misha,misha1}, is that tensorial
spaces, and in particular supergroup manifolds $OSp(N|2n)$, should
play more fundamental role in the construction of the higher spin
field theory than the conventional Minkowski or AdS space--time.

\section{Review of the tensorial particle model}

The action of the $N=1$, $D=4$ superparticle with tensorial
coordinates has the following form
\begin{equation}\label{1}
S=\int\;d\tau\;\lambda_\alpha\lambda_\beta\;(\partial_\tau
X^{\alpha\beta}-i\; \partial_\tau\theta^\alpha\,\theta^\beta)\;,
\end{equation}
where
\begin{equation}\label{2}
X^{\alpha\beta}=X^{\b\a}={1\over
2}\,\gamma^{\alpha\beta}_m\;x^m\,+
\,{1\over 4}\,\gamma^{\alpha\beta}_{mn}\;y^{mn}\,,
\end{equation}
$x^m=\gamma^m_{\a\b}\,X^{\a\b}$ are conventional $D=4$ space--time
coordinates ($m$=0,1,2,3), $y^{mn}=\gamma^{mn}_{\a\b}\,X^{\a\b}$
are antisymmetric tensor coordinates, $\theta^\alpha$ are
anticommuting Majorana spinor coordinates and $\lambda_\alpha$ are
auxiliary commuting Majorana spinor variables
($\alpha=1,\cdots,4$). The physical meaning of the tensorial
coordinates $y^{mn}$ is to provide the superparticle, upon
qunatization, with infinitely many integer and half--integer
spinning degrees of freedom \cite{exotic}. So the generalized
superspace $(X^{\a\b}, \, \theta^\alpha)$ can be regarded as an
extended `higher-spin' superspace which, as we shall see below,
possesses an $OSp(1|8)$ symmetry (see also \cite{misha,misha1} and
references therein) and can be realized as a coset superspace
${{OSp(1|8)}\over{GL(4)\times\!\!\!\!\supset SK}}$, where
$GL(4)\times\!\!\!\!\!\!\supset SK$ is a semidirect product of the
general linear group $GL(4)$ and a generalized super Poincare
group $SK$ which are subgroups of $OSp(1|8)$.

The action is manifestly invariant under $SO(1,3)$ Lorentz rotations
\be\label{2a}
\delta \lambda_\a=l_{\a}^{~~\b}\l_\b\,,\quad \delta
\theta_\a=l_{\a}^{~~\b}\theta_\b\,,\quad\delta
X^{\alpha\beta}=-X^{\alpha\g}\,l_{\g}^{~~\b}-X^{\g\b}\,l_{\g}^{~~\a}\,,
\ee
constant translations in the space of $X^{\alpha\beta}$
\begin{equation}\label{3}
\delta X^{\alpha\beta}=a^{\alpha\beta}
\end{equation}
and under global $N=1$, $D=4$ supersymmetry transformations
\begin{equation}\label{4}
\delta\theta^{\alpha}=\epsilon^{\alpha}, \quad \delta X^{\alpha\beta}=-{i\over
2}(\epsilon^{\alpha}\theta^\beta+\epsilon^{\beta}\theta^\alpha),
\end{equation}
the corresponding superalgebra being
\begin{equation}\label{5}
\{Q_\alpha,\,Q_\beta\}=P_{\alpha\beta}=-2\gamma_{\alpha\beta}^m\;P_m\,+
\,\gamma^{mn}_{\alpha\beta}\;Z_{mn}\,, \quad [P_{\a\b},\,
P_{\g\d}]=[P_{\a\b},\,Q_\g]=0\,,
\end{equation}
where $Q_\alpha$ are supersymmetry generators, $P_m$ are $D=4$
translations and $Z_{mn}=-Z_{nm}$ are tensorial charges generating
the translations along $y^{mn}$.

 From eq. \p{1} it follows that the particle momentum in the $D=4$
space is \be\label{5.1} P_m={1\over 2}\l\gamma_m\l\, \quad
\Rightarrow \quad P_mP^m = 0, \ee
 and hence the particle is
massless. The momenta conjugate to $y^{mn}$ are also constrained
to be the bilinear combinations of the commuting spinors
\be\label{5.2} Z_{mn}={1\over 4}\l\gamma_{mn}\l\,. \ee

The constraints \p{5.1}, \p{5.2} together with the constraints on
the momenta of $\lambda$
\be\label{5.3}
P^\alpha=0,
\ee
and constraints on the momenta conjugate to $\theta$
\be\label{5.4}
\pi_\a=\lambda_\a\,\l_\b\,\theta^\b
\ee
imply that in the phase space the superparticle \p{1} has only
eight bosonic and one fermionic degrees of freedom (see
\cite{exotic} for the details).

The constraints \p{5.1}--\p{5.4} are a mixture of the first and
second class constraints. As has been mentioned in the
Introduction, one of the ways to handle these constraints and to
perform quantization considered in \cite{exotic} is to convert all
the constraints into the first class by introducing auxiliary
variables. This can be done \cite{misha} by adding to the action
\p{1} the terms of the form
\begin{equation}\label{add}
S_{add}=\int d\tau [\lambda_\alpha\partial_\tau
P^\alpha-i\chi(\partial_\tau\chi-\lambda_\a\partial_\tau\theta^\a)],
\end{equation}
where $\chi(\tau)$ is a Grassmann odd variable and $\lambda_\a$ is
regarded as a momentum conjugate to $P^\alpha$ which now is
unconstrained.

In what follows we shall follow an alternative way. The
constraints can be implicitly solved in terms of independent
($8_b,1_f$) supertwistor degrees of freedom by rewriting the
action \p{1} in the following supertwistor form
\begin{equation}\label{6}
S=-\int\;d \tau\;{\cal Z}^a\,\partial_\tau{\cal Z}_a\,,
\end{equation}
where
\begin{equation}\label{7}
{\cal Z}_a=(\lambda_\alpha,\,\mu^\beta,\, \chi)
\end{equation}
is the supertwistor and
\begin{equation}\label{8}
\mu^\alpha=(X^{\a\beta}-{i\over 2}\theta^\a\,\theta^\beta)\lambda_\beta, \quad
\chi=\theta^\alpha\lambda_\alpha\,
\end{equation}
(small Latin letters from the beginning of the alphabet stand for the
supertwistor indices).

 The supertwistor index is raised and lowered by the $OSp\,(1|8)$
invariant matrix
\begin{equation}\label{9}
 C^{ab}=\left(
\begin{array}{ccc}
 0& \delta^{\a}_{\b} & 0\\
-\delta^{\b}_{\a}& 0 & 0\\
0& 0 & -i
\end{array}
\right)\,.
\end{equation}
${\cal Z}^a$ is the ``conjugate" supertwistor
\be\label{10}
{\cal Z}^a=C^{ab}{\cal Z}_b=(\mu^\alpha,\,-\lambda_\beta,\,
-i\chi)\,
\ee
so that the bilinear form ${\cal Z}^a_1{\cal Z}_{2a}$ is $OSp\,(1|8)$ invariant
and so does the action \p{6}. This implies that also the action
\p{1} possesses the $OSp\,(1|8)$ symmetry, and we should only find
corresponding $OSp\,(1|8)$ variations of the coordinates
$X^{\alpha\beta}$ and $\theta^\a$.

\section{The $OSp\,(1|8)$ transformations}

The supertwistors \p{7} and \p{10} transform under a fundamental
linear representation of $OSp\,(1|8)$. The infinitesimal
$OSp\,(1|8)$ transformations
 \be\label{12}
 \delta {\cal
Z}^a={\Xi}^{ab}{\cal Z}_b\,
 \ee are given by
symmetric matrices
 \be\label{13}
 {\Xi}^{ab}=\left( \ba{ccc}
 a^{\a\b} & g^{~\a}_{\b} & -i\epsilon^\a\\
g^{~\b}_{\a} & k_{\a\b} & -i\kappa_\a \\
-i\epsilon^\beta &-i\kappa_\beta  & 0\\
\ea
\right)
\ee
The corresponding symmetric matrix of the $OSp\,(1|8)$ generators
is
\be\label{14}
{T}_{ab}=\left(
\ba{ccc}
P_{\a\b} & G_{\a}^{~\b}& Q_\a\\
G^{~\a}_{\b} & K^{\a\b} & S^\a\\
Q_\b  & S^\beta & 0
\ea
\right)\,.
\ee
In terms of the canonical supertwistor variables ${\cal Z}_{a}$
\p{7}, \p{8} satisfying the Poisson bracket relations $[{\cal
Z}_{a},\,{\cal Z}_{b}]=C_{ab}$, the $OSp(1|8)$ generators are
realized as follows
\be\label{14.1}
{T}_{ab}={\cal Z}_a{\cal Z}_b = \left(
\ba{ccc}
\l_{\a}\,\l_{\b} & \l_\a\,\mu^\b& \l_\a\,\chi\\
\mu^{\a}\,\l_{\b} & \mu^{\a}\,\mu^{\b} & \mu^\a\,\chi \\
\l_\b\,\chi  & \mu^\beta\,\chi & 0
\ea
\right)\,.
\ee

The components of \p{14} satisfy the $osp(1|8)$ superalgebra
$$
[P,P]=0\,,\quad [K,K]=0,
$$
\be\label{15b}
 [P,K]\sim G\,, \quad [G,G]\sim
G\,,
\quad [G,P]\sim P\,, \quad [G,K]\sim K\,,
\ee
\be\label{15c}
\quad [G,Q]\sim Q\,, \quad [G,S] \sim S\,,
\quad
\{Q,Q\}\sim P\,, \quad \{S,S\}\sim K\,,\quad \{S,Q\}\sim G\,,
\ee
where
\be\label{15}
T_{\hat\a\hat\b}=
\left(
\ba{cc} P_{\a\b} & G_{\a}^{~\b} \\
G_{\b}^{~\a} & K^{\a\b}
\ea
\right), \quad (\hat \a,\hat \b=1,\cdots , 8)
\ee
are the generators of the $Sp(8)$ subgroup of $OSp\,(1|8)$.

The components of $T_{\hat\a\hat\b}$ are the $GL(4)$ general
linear transformations $G_{\a}^{~\b}$ , the symmetric commuting
tensorial charges $P_{\a\b}$ which include the $D=4$ translations
$P_m$ and the tensorial boosts $Z_{mn}$,  and which together with
the fermionic charges $Q_\a$ generate the supersymmetry algebra
\p{5}, and `generalized conformal' boosts
\be\label{15a}
K^{\a\b}=K^{\b\a}=\gamma^{\alpha\beta}_m\;K^m\,+\,\gamma_{mn}^{\alpha\beta}\;K^{mn}\,,
\ee
where $K^m$ are the standard special conformal generators and
$K^{mn}=-K^{nm}$ are their tensorial counterparts.

 The $SO(1,3)$ spinor indices can be raised and lowered with
antisymmetric charge conjugation matrices $C^{\a\b}$ and
$C_{\a\b}$.
\begin{equation}\label{11}
\l^\a=C^{\a\b}\l_\beta \,\,,\quad \mu_\a=-C_{\a\b}\,\mu^\beta\,,\quad
C^{\a\g}C_{\g\b}=-\delta_\b^{\a} \,.
\ee
Let us consider the matrix $G_{\a\b}=G^{~\g}_\a\,C_{\g\b}$
obtained by lowering the index of the GL(4) matrix.
 The symmetric part of $G_{\a\b}$ generates an $Sp(4)\sim
SO(2,3)$ subgroup of $Sp(8)$ which contains the $D=4$ Lorentz
group generated by $L_{mn}$ and vector charge generators $Z_m$
which can be regarded as $AdS_4$ boosts
\be\label{16}
G_{\{\a\b\}}\equiv M_{\a\b}=\gamma_{\alpha\beta}^m\;Z_m\,+
\,\gamma^{mn}_{\alpha\beta}\;L_{mn}\,. \ee The antisymmetric part
of $G_{\a\b}$ contains dilatation $D$, a $U(1)$ generator U and a
3--form charge $Z_{mnp}$
\be\label{17}
G_{[\a\b]}={1\over 2}
C_{\alpha\beta}\,D+\g^5_{\a\b}\,U+\gamma_{\alpha\beta}^{mnp}\;Z_{mnp}\,.
\ee
$OSp\,(1|8)$ can be regarded as a generalized $N=1$, $D=4$ superconformal
group with $S^\a$ being superconformal generators
\be\label{18}
\{S^\a,S^\b\}=K^{\a\b}\,.
\ee
Note that $S^\a$ and $K^{\a\b}$ generate a second copy of the
generalized super Poincare algebra (SK) similar to \p{5}.

$OSp\,(1|8)$ contains the conformal group $SO(2,4)$ as a subgroup
generated by $P_m$, $L_{mn}$, $D$ and $K^m$, but the
superconformal group $SU(2,2|1)$ is not a subgroup of $OSp\,(1|8)$
because of the reasons explained in \cite{conflat}. $SU(2,2|1)$ is
a subgroup of a larger simple supergroup $OSp\,(2|8)$. We shall
consider a generalization of the superparticle model \p{1}, which
is $OSp\,(2|8)$ invariant, in the next Section.

Using the variation law \p{12}, \p{13} and the particular form of
the supertwistor components \p{7}--\p{10} we can easily get the
$OSp\,(1|8)$ variation of $\lambda_\a$, $\theta^\a$ and $X^{\a\b}$
\be\label{18a}
\delta\lambda_\a=g_\a^{~\b}\lambda_\b-k_{\a\b}(X^{\b\g}-{i\over
2}\theta^\b\theta^\g)\lambda_{\gamma}-i\kappa_\a\theta^\b\lambda_\b\,,
\ee
\be\label{19}
\delta
\theta^\a=\e^\a-\theta^\b\,g_\beta^{~\a}
+\theta^\b\,k_{\b\g}X^{\g\a}-
\kappa_\b(X^{\b\a}-{i\over
2}\theta^\b\theta^\a)-i\theta^\a\theta^\b\kappa_\b\,,
\ee
$$
\delta X^{\a\b} =
[a^{\a\b}-{i\over
2}(\epsilon^{\alpha}\theta^\beta+\epsilon^{\beta}\theta^\alpha)]-
(X^{\g\b}g_\g^{~\a}+X^{\g\a}g_\g^{~\b})+X^{\a\g}k_{\g\d}X^{\d\b}-
$$
\be\label{20}
 - {i\over 2}\theta^\a X^{\b\g}\kappa_\g-{i\over 2}\theta^\b
X^{\a\g}\kappa_\g\,.
\ee
In \p{19} and \p{20} the first terms are $N=1$ supersymmetry
transformations \p{3}, \p{4} in the $D=4$ superspace extended with
the tensorial directions, the second terms contain $SO(1,3)$
Lorentz rotations \p{16} (when $g^{\a\b}= l^{mn}\,\g_{mn}^{\a\b}$)
and dilatations
\p{17} (when
$g^{\a\b}={1\over 2}C^{\a\b}\,\phi$), and the remaining terms are generalized
conformal and superconformal boosts.

The action \p{1}, as well as the constraints and the equations of
motion which it produces, are invariant under the $OSp\,(1|8)$
transformations \p{18a}--\p{20}. Note that under \p{18a}--\p{20}
the one forms
\be \label{1F}
\Pi^{\a\b}=dX^{\alpha\beta}-{i\over 2}\;
(d\theta^\alpha\,\theta^\beta+ d\theta^\beta\,\theta^\alpha),
\qquad E^\a=d\theta^\a
\ee
 transform as follows
\be\label{20.1}
\delta \Pi^{\a\b}= \Pi^{\a'\b}\,g_{\a'}^{~~\a}(X,\theta)\,+\,
 \Pi^{\a\b'}\,g_{\b'}^{~~\b}(X,\theta)\,, \quad
\ee
where $g_{\a'}^{~\a}(X,\theta)$ are the infinitesimal
$OSp(1|8)$ transformations nonlinearly realized on the coset
superspace ${{OSp(1|8)}\over{GL(4)\times\!\!\!\!\supset SK}}$ in
terms of $GL(4)$ matrices. $SK$ stands for the copy of the
generalized super Poincare subgroup of $OSp(1|8)$ generated by
$S_\a$ and $K_{\a\b}$ \p{18}. We have thus demonstrated that the
generalized superspace $(X^{\a\b},\,\theta^\a)$ is the coset
superspace ${{OSp(1|8)}\over{GL(4)\times\!\!\!\!\supset SK}}$.

Note that the finite $OSp\,(1|8)$ variations of $\lambda_\a$ and
$\Pi^{\a\b}$ are
\be\label{20.2}
\hat\lambda_\a=G^{-1\a'}_\a(X,\theta)\,\lambda_{\a'}, \quad \hat
\Pi^{\a\b}= \Pi^{\a'\b'}\,G_{\a'}^{~~\a}(X,\theta)\,
G_{\b'}^{~~\b}(X,\theta)\,,
\ee
from which one immediately sees that the action \p{1} is $OSp\,(1|8)$
invariant.

\section{The $OSp(2|8)$ invariant model}

As we have mentioned, a conventional superconformal group $SU(2,2|1)$ is not a
subgroup of the supergroup $OSp(1|8)$ but is a subgroup of $OSp(2|8)$. As it
has been found in \cite{exotic}, there exists a generalization of the action
\p{1} which is $OSp(2|8)$ invariant. It has the following form
\begin{equation}\label{21}
S=\int\;d\tau\;\lambda_\alpha\lambda_\beta\;(\partial_\tau
X^{\alpha\beta}-{i\over 4}\;
\partial_\tau\theta\gamma^m\theta\gamma_m^{\a\b}
+{i\over 8}\,a\,\partial_\tau\theta\gamma^{mn}\theta\gamma_{mn}^{\a\b})\;,
\end{equation}
where $0\leq a\leq 1$ is a numerical parameter and
\be\label{211}
X^{\alpha\beta}=X^{\b\a}={1\over
2}\,\gamma^{\alpha\beta}_m\;x^m\,+ \,{a\over
4}\,\gamma^{\alpha\beta}_{mn}\;y^{mn}\,.
\ee

When $a=1$ eq. \p{21} reduces to \p{1} due to the Fierz identity
\be\label{22}
C^{\a\{\b}C^{\g\}\d}={1\over 4}\gamma^{m\a\d}\gamma_m^{\b\g}-{1\over
8}\gamma^{mn\a\d}\gamma_{mn}^{\b\g}\,.
\ee
Using other Fierz identities the action \p{21} can be written as
\begin{equation}\label{23}
S=\int\;d\tau\;\lambda_\alpha\lambda_\beta\;\left[\partial_\tau
X^{\alpha\beta}-{{i(1+a)}\over 2}\;
\partial_\tau\theta^\a\theta^\b
+{{i(1-a)}\over
2}\,(\partial_\tau\theta\gamma^5)^\a(\theta\gamma^{5})^\b\right]\;,
\end{equation}
or in a manifestly invariant $OSp(2|8)$ supertwistor form as
\begin{equation}\label{24}
S=-\int\;d \tau\;{\cal Z}^{\hat a}\,\partial_\tau{\cal Z}_{\hat
a}\,,
\end{equation}
where
\begin{equation}\label{25}
{\cal Z}_{\hat a}=(\lambda_\alpha,\,\mu^\beta,\, \chi^i), \qquad
i=1,2
\end{equation}
and
\bea\label{26}
&\mu^\alpha=\left[X^{\a\beta}-{{i(1+a)}\over
4}\;\theta^\a\,\theta^\beta-{{i(1-a)}\over 4}\;
(\theta\gamma^5)^\a\,(\theta\gamma^5)^\beta\right]\,\lambda_\beta,
\nonumber\\
&\chi^1=\sqrt{{1+a}\over 2}\;\theta^\alpha\lambda_\alpha\,\qquad
\chi^2=i\sqrt{{1-a}\over 2}\;\theta\gamma^5\lambda\,.
\eea
 The supertwistor index is raised and lowered by the $OSp\,(2|8)$
invariant matrix
\begin{equation}\label{27}
 C^{\hat a\hat b}=\left(
\begin{array}{ccc}
 0& \d_{\a}^{\b} & 0\\
-\d^{\a}_{\b}& 0 & 0\\
0& 0 & -i\delta^{ij}
\end{array}
\right)\,.
\end{equation}

As in the previous Section starting from linear $OSp\,(2|8)$ transformations of
the supertwistor \p{25} generated by the matrices
\be\label{28}
{T}_{\hat a \hat b}=\left(
\ba{ccc}
P_{\a\b} & G_{\a}^{~\b}& Q^j_\a\\
G^{~\a}_{\b} & K^{\a\b} & S^{j\a}\\
Q^i_\b  & S^{i\beta} & \epsilon^{ij}
\ea
\right)\,.
\ee
one can find corresponding $OSp\,(2|8)$ variations of $X^{\a\b}$
and
$\theta^\a$.

\section{Quantization and higher spin field equations of motion}
The quantization of the superparticle models described by the
actions \p{1}, \p{21} and \p{23} has been considered in detail in
\cite{exotic} so we shall just present main results which are a
direct consequence of the quantization of the free supertwistor
models \p{6} and \p{24}. In particular, for the values of the
parameter $a\not = 0,1$ the action \p{24} is of the first order
form with $\l^\a$ and  $\chi$ being the coordinates and $\mu_\a$
and $2i\bar\chi$ being the corresponding conjugate momenta. Upon
quantization the Poisson brackets of the canonical variables
\be\label{290}
[\mu_\a,\,\l^\b]_P=\delta^\a_\b,\qquad
\{\bar\chi,\,\chi\}_P=-{i\over 2}
\ee
are replaced with the (anti)commutators according to the rule
$[~,\,~]_P~\rightarrow~ i[~,\,~]$ and $\{~,\,~\}_P~\rightarrow
-i\{~,\,~\}$, and the quantum states of the superparticle are
described by a wave function  of $\lambda_\a$ and
$\chi=\chi^1+i\chi^2$
\be\label{29}
\Phi(\lambda_\alpha,\,\chi)=\sum_{n=0}^\infty\,
\lambda_{\a_1}\cdots\lambda_{\a_n}(\phi^{\a_1\cdots
\a_n}+i\chi\,\psi^{\a_1\cdots \a_n})\,.
\ee
For further analysis it is useful to rewrite the wave function \p{29} in a
two--component Weyl representation of the spinor $\l^{\a}$
\be\label{30}
\l^{\a}=(\l^A,\,\bar\l^{\dot A}),  \qquad (A,\dot A=1,2)
\ee
where $\l^A$ and $\bar\l^{\dot A}$ are complex conjugate Weyl spinors. We thus
have
\bea\label{31}
\Phi(\lambda_A,\,\bar\l_{\dot A}\,,\chi)&=
\sum_{n=0}^\infty\,
\lambda_{A_1}\cdots\lambda_{A_n}[\phi^{A_1\cdots
A_n}(\lambda_A\bar\lambda_{\dot A})+i\chi\,\psi^{A_1\cdots
A_n}(\lambda_A\bar\lambda_{\dot A})]\,\nn\\
& +\,\sum_{n=0}^\infty\, \bar\lambda_{\dot
A_1}\cdots\bar\lambda_{\dot A_n}[\bar\phi^{\dot A_1\cdots \dot
A_n}(\l_A\bar\lambda_{\dot A})+i\chi\,\bar\psi^{\dot A_1\cdots
\dot A_n}(\lambda_A\bar\lambda_{\dot A})]\,,
\eea
where $\phi(\l_A\bar\lambda_{\dot A})$ and $\psi(\l_A\bar\lambda_{\dot A})$
depend on the product of the complex conjugate spinors, which because of
\p{5.1} is equal to the superparticle momentum
\be\label{32}
P_{A\dot A}\equiv \sigma^m_{A\dot A}P_m=\l_A\bar\lambda_{\dot A},
\ee
$\sigma^m_{A\dot A}$ being Pauli matrices. Therefore, the functions \p{29} and
\p{31} describe quantum states of the superparticle in the momentum
representation. This has been obtained in \cite{exotic}.

In the case $a=1$ there is only one real fermionic variable $\chi$
(see eq. \p{26}) which becomes a Clifford variable ($(\chi)^2=1$)
when one applies the standard Dirac quantization procedure and
solves the second class constraint relating $\chi$ with its
conjugate momentum. As a result the wave functions \p{29} and
\p{31} become Clifford `superfields' \cite{exotic}.

Finally, when $a=0$ in the action \p{23} there is an additional
first class constraint
\be\label{sv}
\mu^A\l_A-\bar\mu^{\dot A}\bar\l_{\dot A}+2i\bar\chi\chi=0.
\ee
This constraint implies a helicity condition on the wave function
\p{31}
\be\label{hc}
(\l_A{\partial\over{\partial\l_A}}-\bar\l_{\dot
A}{\partial\over{\partial\bar\l_{\dot
A}}}+\chi{\partial\over{\partial\chi }}-s)\Phi(\l,\chi)=0\,,
\ee
where $s$ is an integer constant which appears because of a
quantum ordering ambiguity in \p{sv} and takes several fixed
values \cite{eisen,mp}. As a result, when $a=0$ the spectrum is
restricted to the one of a massless (anti)chiral $N=1,\,D=4$
supermultiplet with $s$ being the corresponding superhelicity. On
the contrary, when $a\not = 0$ the helicity constraint is absent
and one has the infinite tower of massless fields of arbitrary
integer and half integer helicity which we shall now show to obey
the Vasiliev unfolded higher spin field equations.

 For this we should pass from the twistor wave functions \p{31}
to $D=4$ coordinate wave functions which satisfy standard
equations of motion of corresponding massless spin fields.  This
is achieved by performing a Fourier transformation using a
prescription analogous to one proposed in \cite{eisen,mp} (see
also \cite{igor,fezi} for relevant formulations in terms of
Lorentz harmonics and `index' spinors). Before presenting a
general formula, let us consider several simple examples.

\subsubsection*{s=0}
The spin zero quantum state of the tensorial superparticle is described by the
$n=0$ component $\phi_0(\l_A\bar\lambda_{\dot A})$ of \p{31}. To get its
coordinate representation we just multiply $\phi_0(\l_A\bar\lambda_{\dot A})$
by $\exp{(ix^mP_m)}$ (where $P_m$ is defined in \p{32}) and integrate over
$\l_A$ and $\bar\lambda_{\dot A}$
\be\label{33}
\phi(x)={1\over {2\pi}}\int\,d^2\lambda \,d^2\bar\lambda\,
 e^{ix^mP_m}\,\phi_0(\l_A\bar\lambda_{\dot A}), \quad
 P_m=2\lambda\sigma_m\bar\l\,.
\ee
Note that integration involves four independent components of $\l_A$ and
$\bar\lambda_{\dot A}$ three of which are associated with the components of the
lightlike momentum $P_m$ and the fourth one is the phase
\be\label{33.1}
\lambda_A=e^{i\varphi}\lambda_{0A}\,, \qquad 0\leq\varphi<2\pi\,,
\ee
under which $P_m$ of eq. \p{32} is
invariant. So the integration over $\varphi$ gives a factor of $2\pi$ which is
canceled in \p{33} by the normalization constant. Because of the relation
\p{32} the wave function \p{33} satisfies the Klein--Gordon equation
\be\label{34}
{\partial^2\over{\partial x^m\partial x_m}}\phi(x)=0\,.
\ee
This equation just reflects the fact that in the classical model
there is a first class constraint $P_mP^m=0$ \p{5.1} which ensures
the quantum states of the superparticle to be massless.

\subsubsection*{s=1/2}
The spin $s={1\over 2}$ states come from the $n=1$ component of
$\phi$ or from the $n=0$ component of $\psi$ (eq. \p{31}). In the
latter case the spin 1/2 state is a superpartner of the scalar
state discussed above, since $\chi=\theta^A\lambda_A$. Consider,
for instance, the function $\bar\lambda_{\dot A}\,\bar\phi^{\dot
A}(\l\bar\lambda)$. To convert it into the coordinate spinorial
wave function we multiply it by $\exp{(ix^mP_m)}$ and
$\lambda_{A}$ (so that again $\lambda$ and $\bar\lambda$ appear in
pairs) and perform the integration as in eq. \p{33}
\be\label{35}
\psi_{A}(x)={1\over{2\pi}}\,\int\,d^2\lambda
\,d^2\bar\lambda\,e^{ix^mP_m}\,\, \lambda_A\, \bar\lambda_{\dot
A}\,\bar\phi^{\dot A}(\l\bar\lambda)\,.
\ee
By construction, in view of \p{32}, the spinorial function \p{35} satisfies the
Weyl equation for a massless spin 1/2 field
\be\label{36}
\sigma^m_{A\dot A}\,\partial_m\,\psi^{ A}(x)=0.
\ee
Note that if we did not multiply $\bar\lambda_{\dot
A}\,\bar\phi^{\dot A}(\l\bar\lambda)$ by $\lambda_{A}$ the
integrated expression would depend on the phase factor \p{33.1}
and the integral $\int\, d\varphi\,e^{i\varphi}$ would vanish.
This is a generic situation. The integral over $\lambda$ and
$\bar\l$ vanishes unless the number of $\lambda$ and $\bar\l$ in
the integrated expression match.

At $a=0$ the chiral $N=1$, $D=4$ superfield which describes a
scalar $N=1$ supermultiplet of spin 0 and spin 1/2 states is
extracted from \p{31} as follows
\be\label{36.1}
\Phi(x,\theta)={1\over{2\pi}}\,\int\,d^2\lambda \,d^2\bar\lambda\,
e^{i(x^m-i\theta\sigma^m\bar\theta)P_m}\,\,
\Phi(\lambda,\,\bar\l\,,\chi)\,, \quad \chi=\lambda_A\,\theta^A\,.
\ee

\subsubsection*{s=1}
Using the same procedure as above one gets from \p{31} the self--dual and
anti--self--dual field strengths of an abelian gauge field
$$
F_{AB}(x)={1\over{2\pi}}\,\int\,d^2\lambda
\,d^2\bar\lambda\,e^{ix^mP_m}\,\, \lambda_{A}\lambda_B\,
\Phi(\lambda,\,\bar\l\,,\chi)|_{\chi=0}\,=
$$
\be\label{37}
={1\over{2\pi}}\,\int\,d^2\lambda \,d^2\bar\lambda\,e^{ix^mP_m}\,\,
\lambda_{A}\lambda_B\, \bar\l_{\dot A_1}\bar\l_{\dot
A_2}\,\bar\phi^{\dot A_1{\dot A_2}},
\ee
$$
F_{\dot A\dot B}(x)={1\over{2\pi}}\,\int\,d^2\lambda
\,d^2\bar\lambda\,e^{ix^mP_m}\,\,\bar\lambda_{\dot
A}\bar\lambda_{\dot B}\, \Phi(\lambda,\,\bar\l\,,\chi)|_{\chi=0}=
$$
\be\label{38}
= {1\over{2\pi}}\,\int\,d^2\lambda
\,d^2\bar\lambda\,e^{ix^mP_m}\,\, \bar\lambda_{\dot
A}\bar\lambda_{\dot B}\, \l_{A_1}\l_{A_2}\,\phi^{A_1{A_2}}\,,
\ee
$F_{AB}(x)$ and $F_{\dot A\dot B}(x)$ satisfy the Bianchi
identities and the Maxwell equations written in the following form
\be\label{39}
\sigma^m_{\dot AA}\partial_m\,F^{AB}(x)=0\, , \qquad
\sigma^m_{A\dot A}\partial_m\,F^{\dot A\dot B}(x)=0\,.
\ee

\subsection{Higher spins and Vasiliev's unfolded dynamics}
We are now in a position to write a general expression which
relates the wave function \p{31} with a generating function
$C(\bar a, b, \chi|x)$ which satisfies conformal field equations
describing an `unfolded' higher spin dynamics in flat $D=4$
space--time \cite{vas,misha,misha1}. $C(\bar a, b, \chi|x)$
depends on auxiliary spinor variables $\bar a^{\dot A}$, $b^A$ and
$\chi$ satisfying star product commutation relations (actually,
$\bar a^{\dot A}$ and $b^A$ are creation operators acting in a
Fock space of higher spin states \cite{misha})
\be\label{40}
[a_A,\,b^B]=\delta_A^{~B}\,,\qquad [\bar a_{\dot A},\,\bar b^{\dot
B}]=\delta_{\dot A}^{~\dot B}\,,
\qquad \{\chi\,,\bar\chi\}=1\,
\ee
and is expressed in terms of $\Phi(\l,\bar\l,\chi)$ (eq. \p{31})
as follows
$$
C(\bar a, b,\chi|x) \equiv
\sum_{m,n=0}^\infty \,\sum_{k=0}^1\,{1\over{m!n!}}\bar a^{\dot A_1}
\cdots\bar a^{\dot A_m}\,b^{
A_1}\cdots b^{A_n}\,(\chi)^k\,C_{A_1\cdots A_n\dot A_1\cdots\dot
A_m,\,k}\,(x)
$$
\be\label{41}
 =\sum_{m,n=0}^\infty \,{1\over{2\pi m!n!}}\,\int\,d^2\l\,d^2\bar\l\,
\,(\bar a^{\dot A}\bar\lambda_{\dot A})^m\,(b^{
A}\lambda_{ A})^n\,e^{ix^{k}\lambda\sigma_k\bar\lambda}
\,\Phi(\l,\bar\l,\chi)\,.
\ee
By construction \p{41} satisfies the equations of unfolded higher
spin dynamics (see \cite{vas,misha} for details)
\be\label{42}
i{{\partial^2}\over{\partial b^A\,\partial\bar a^{\dot A}}}C(\bar
a, b,\chi|x)=\sigma^m_{A\dot A}{\partial\over{\partial x^m}}C(\bar
a, b,\chi|x)\,.
\ee
The equation \p{42} reproduces dynamical equations of motion of
massless higher spin fields described by holomorphic
$C(0,b,\chi|x)$ and anti--holomorphic $C(\bar a,0,\chi|x)$ components of
\p{41} and expresses non--holomorphic components, which are auxiliary
fields, as space--time derivatives of the (anti)holomorphic
physical fields.

As has been discussed in detail in \cite{misha} the unfolded
system of equations \p{42} is invariant under an infinite
dimensional higher spin algebra $hu(1,1|8)$ which mixes different
higher spin components of the generating function
\p{41}. $hu(1,1|8)$ contains a finite dimensional superconformal subalgebra
$osp(2|8)$. This $osp(2|8)$ superconformal symmetry of the higher
spin field equations is a manifestation of the $osp(2|8)$
invariance of the classical action \p{21} whose quantization
produced the massless higher spin states.

We have thus found a direct relationship of the quantized
superparticle in the generalized tensorial coordinate superspace
with the unfolded classical dynamics of free higher spin fields by
Vasiliev.

\section{Tensorial superparticles and $AdS$ superspaces. $OSp(1|2n)$ Cartan forms}

In the previous sections we have discussed the dynamics of a
superparticle in a generalized superspace parametrized by bosonic
tensorial coordinates $X^{\a\b}$ and Grassmann--odd spinor
coordinates $\theta^{\a}$ having a flat $N=1$, $D=4$ superspace
$(x^m=\gamma^m_{\a\b}\,X^{\a\b}$, $\theta^{\a}$) as a subspace.

In order to obtain  higher spin fields in
 an $AdS_4$ background
we should quantize the superparticle propagating on a tensorial
extension of the $AdS_4$ superspace. The reason for this is the
following. As we have seen in the previous section, the
quantization of the $OSp(N|8)$ ($N=1,2$) invariant superparticle
in the flat tensorial superspace
 leads to the equations describing free
massless higher spin fields.
 The presence of the tensorial coordinates is
crucial for the possibility of obtaining the infinite tower of the
higher spin states. For the description of higher spin fields on
$AdS_4$ we should therefore find the corresponding generalization of the
flat tensorial space to a space which along with the coordinates
corresponding to $AdS_4$  contains additional tensorial
coordinates.

We shall now analyze whether there exists a generalized superspace
which contains in some form an $AdS_4$ superspace
${{OSp(1|4)}\over{SO(1,3)}}$. An assumption has been made
\cite{preit} (see also \cite{misha}) that this should be a
supergroup manifold $OSp(1|4)$ whose appropriate truncation is
known to result in the flat tensorial superspace. Note that both
the $N=1$, $D=4$ super Poincar\'e group extended with the
tensorial charge \p{5} and the supergroup $OSp(1|4)$ are subgroups
of $OSp(1|8)$ and $OSp(2 |8)$ which are symmetries of the
tensorial superparticle action \p{1} and \p{21}, respectively.

We start the analysis by observing that the generators of the
$AdS_4$ boosts can be singled out from the generators of the four
dimensional conformal group $SO(2,4)$ by taking a linear
combination of the generators of Poincar\'e translations $P_m$ and
conformal boosts $K_m$, namely ${\cal P}_m=P_m+K_m$. The $AdS_4$
manifold can be considered as a coset space
$SO(2,4)\over{(SO(1,3)\times D) \times\!\!\!\!\supset K}$
parametrized by the  coset element $e^{{\cal P}_m\,x^m}$.

Analogously, for the case of tensorial extension of $AdS_4$ space
let us consider the generators
\be\label{6.1}
 {\cal P}_{\a\b}= P_{\a\b}+K_{\a\b}, \quad [{\cal P},{\cal P}]\sim
M,
 \quad [{\cal P},M]\sim {\cal P},
\ee
 where $M_{\a\b}$ stand for the generators of $Sp(4)$  \p{16}.
One can see that ${\cal P}_{\a\b}$ and $M_{\a\b}$ form the algebra
of $Sp_L(4)\times Sp_R(4)$, where $Sp(4)$ generated by $M_{\a\b}$
is the diagonal subgroup. The generators of $Sp_L(4)$ are
$M_L=P+K+M$ and that of $Sp_R(4$) are $M_R=P+K-M$. Thus the
extended tensorial $AdS_4$ is a coset (group) manifold
\be\label{6.2}
 Sp(4)\sim {{Sp_L(4)\times Sp_R(4)}\over{Sp\,(4)}}.
\ee
 $Sp\,(4)$ is a diagonal subgroup of $Sp_L(4)\times Sp_R(4)$,
 which is a subgroup of $Sp(8)$
and this manifold can be realized as a coset space $
{{Sp(8)}\over{GL(4) \times\!\!\!\!\supset K}}$ with the coset
element $e^{(P+K)_{\a\b}\,X^{\a\b}}$.

 Let us note  also that, as we have discussed in Section 3,
  the flat tensorial space associated with $P_{\a\b}$
can be realized as a formally similar but different coset space $
{{Sp(8)}\over{GL(4) \times\!\!\!\!\supset K}}$ with
$e^{P_{\a\b}\,X^{\a\b}}$ being the coset element.

As to the superspace generalization, to get the tensorial
extension of super $AdS_4$ we should add to the bosonic generators
${\cal P}=P+K$ the fermionic generators
\be\label{6.55}
{\cal  Q}= Q+S, \quad \{{\cal Q},{\cal Q}\}\sim P+K+M=M_L\,.
\ee
One can see that $\cal Q$ and $M_L=P+K+M$ form the superalgebra of
$OSp_L(1|4)$, and the coset superspace in question is
\be\label{6.5}
{{OSp_L(1|4)\times Sp_R(4)}\over{Sp\,(4)}}.
\ee

One should also note that this supercoset is isomorphic to
$OSp(1|4)$. I.e. the simplest tensorial extension of the $N=1$
supersymmetric $AdS_4$ is the supergroup manifold $OSp(1|4)$. Like
the flat tensorial superspace (see the discussion after eq.
\p{20.1}) the tensorial extension of the $AdS_4$ superspace can be
realized as a coset superspace
\be \label{COSS}
{{OSp(1|8)}\over{GL(4)\times\!\!\!\!\!\!\supset SK}}\,,
\ee
both being embedded in a certain way into $OSp(1|8)$.

The next question we would like to address is whether the
supergroup manifold $OSp(1|4)$ is related via a $GL(4)$
transformation to the flat tensorial superspace whose
supervielbeins transform under superconformal $Sp(1|8)$ by induced
$GL(4)$ rotations as has been shown in eqs. \p{1F}--\p{20.2}. We
know from \cite{conflat} that the conventional $AdS_4$ superspace
$OSp(1|4)\over SO(1,3)$ is superconformally flat,  i.e. the
corresponding supervielbeins  can be written in the form
\be\label{6.10}
E^{A \dot A} = e^{\Phi(x,\theta, \bar \theta)} \Pi^{A \dot
A}=e^{\Phi(x,\theta, \bar \theta)} (dX^{A\dot
A}-id\theta^A\bar\theta^{\dot A}+i\theta^A d\bar\theta^{\dot A})
\ee
\be\label{6.11}
E^A = e^{\Phi(x,\theta, \bar \theta) +i W(x,\theta, \bar \theta)}
(d \theta^A + 2i \Pi^{A \dot A}\bar D_{\dot A} \Phi ),
\ee
where $\Phi(x,\theta, \bar \theta)$  and $W(x,\theta, \bar
\theta)$ are superfunctions. We see, for example, that the vector
$AdS_4$ supervielbein \p{6.10} is related to the flat superspace
supervielbein $\Pi^{A\dot A}$ by a Weyl rescaling. For an earlier
detailed discussion of various aspects of superconformal field
theories in the $AdS_4$ superspace and their relation to
superconformal theories in flat superspace see \cite{is}.

In our case of the tensorial superspaces the role of the ``Weyl"
factor is played by the $GL(4)$ matrices (see \p{20.2}). An
analogy with \p{20.2} allows us to assume that the bosonic
supervielbeins $\Omega^{\a\b}$ of $OSp(1|4)$ can be put into the
form similar to \p{20.2} and in this sense are `$GL(4)$ flat'.
Actually, this turns out to be the case for an arbitrary
supergroup manifold $OSp(1|2n)$ as we shall show below.

Let us first consider the case of a bosonic simplectic group
manifold $Sp\,(2n)$ $(n=1,\cdots,\infty)$. We call a manifold
`$GL(2n)$ flat' if its vielbein (or Cartan form) $\omega^{\alpha
\beta}=dX^{\mu\nu}\,\omega^{~~\alpha\,
\beta}_{\mu\nu}(X)$ is flat up to a $GL(2n)$ rotation\footnote{We
should stress that this is a nontrivial property in
that it implies that the Cartan form component matrix
$\omega^{~~\alpha\,\beta}_{\mu\nu}(X)$ is a ``direct product" of
$GL(2n)$ matrix components
$\omega^{~~\alpha\,\beta}_{\mu\nu}(X)={1\over 2}[
G_{\mu}{}^\alpha\,(X) G_{\nu}{}^\beta\,(X)+G_{\mu}{}^\beta\,(X)
G_{\nu}{}^\alpha\,(X)]$, which of course does not take place in
the case of a generic matrix.}
\be  \label{ANB}
\omega^{\alpha \beta}(X) = d X^{\alpha^\prime
\beta^\prime}\;G_{\alpha^\prime}{}^\alpha\,(X)
G_{\beta^\prime}{}^\beta\,(X) .
\ee
The Maurer--Cartan equations for the group manifold $Sp\,(2n)$
have the form
\be  \label{MKB}
d \omega^{\alpha \beta} + \frac{\alpha}{2}\omega^{\alpha \gamma}
\wedge \omega_\gamma{}^\beta =0\,.
\ee
Substituting \p{ANB} into \p{MKB} and solving for
$G_{\alpha^\prime}{}^{\alpha}$ one gets
\be \label{RESH}
{(G_{\alpha^\prime}{}^{\alpha})}^{-1} =
\delta_{\alpha^\prime}{}^{\alpha} +\frac{\alpha}{4}
X_{\alpha^\prime}{}^{\alpha}, \quad
G_{\alpha^\prime}{}^{\alpha}=\delta_{\alpha^\prime}{}^{\alpha}+
\sum_{n=1}^{\infty}\,\left(-{\alpha\over
4}\right)^n\,{(X)^n}_{\alpha^\prime}{}^{\alpha},
 \ee
 where $\alpha$ is a parameter of inverse length dimension, which
 in the case of $Sp(4)$ is proportional to the inverse $AdS_4$
 radius $\alpha={4\over R}$.
Thus the $Sp(2n)$ is $GL(2n)$ flat and it reduces to the flat
tensorial space of Sections 2--5 when $\alpha=0$.

These results are generalized to a super group manifold
$OSp(1|2n)$ whose Maurer--Cartan equations have the form
\be \label{MKS}
d \Omega^{\alpha \beta} + \frac{\alpha}{2}\Omega^{\alpha \gamma} \wedge
\Omega_\gamma{}^\beta = -i E^\alpha \wedge E^\beta
\ee
\be
d E^{\alpha} + \frac{\alpha}{2} E^\beta \wedge
\Omega_\beta{}^\alpha=0.
\ee
After introducing the new Grassmann variable $\Theta^\alpha$
\be \label{SOLS}
  \theta^\alpha = \Theta^{\alpha^\prime}
  {(G_{\alpha^\prime}{}^{\alpha})}^{-1}
P^{-1}(\Theta^2)\,,
\ee
 one can verify that the Maurer--Cartan equations \p{MKS}
  are satisfied by the
following $GL(2n)$ flat ansatz
\be\label{ANS}
\Omega^{\alpha \beta}= \Pi^{\alpha^\prime  \beta^\prime}\,{\cal
G}_{\alpha^\prime}{}^{\alpha} {\cal G} _{\beta^\prime}{}^\beta ,
\quad {\cal
G}_{\alpha^\prime}{}^\alpha=G_{\alpha^\prime}{}^\alpha\,(X)
-\frac{i\alpha}{8}(\Theta_{\alpha^\prime}-2G_{\alpha^\prime}{}^\gamma
\Theta_\gamma ) \Theta^\alpha\,,
\ee
\be \label{SF1}
E^\alpha = P(\Theta^2){\cal D} \Theta^\alpha - \Theta^\alpha {\cal
D} P(\Theta^2), \quad P(\Theta^2) = \sqrt{1
+\frac{i\alpha}{8}\Theta^\beta \Theta_\beta}\,
\ee
where ${\cal D}=\delta_\a^\b\, d + {\alpha\over
2}\omega_\a^{~\b}(X)$ is the covariant derivative,
$G_{\alpha^\prime}{}^\alpha(X)$ has been defined in  \p{RESH} and
$\Pi^{\alpha
\beta}=dX^{\a\b}-{i\over 2}(d\theta^\a\theta^\b+d\theta^\b\theta^\a)$ being
the flat superform \p{1F}. The bosonic Cartan form (\ref{SF1}) is
bilinear in  the Grassmann coordinates $\Theta$ \p{SOLS}, as has
been found in \cite{preit}
\be\label{6.16}
\Omega^{\alpha \beta} = \omega^{\alpha \beta}(X) +
\frac{i}{2}(\Theta^\alpha {\cal D} \Theta^\beta +\Theta^\beta
{\cal D} \Theta^\alpha ),
\ee
which is an indirect check of the $GL(2n)$ flat ansatz.

\section{Quantization of the superparticle on $OSp(1|4)$ and higher
spin fields on the $AdS_4$ superbackground}

 As it could be
expected from the discussion of the previous section the
quantization of the superparticle propagating on the $OSp(1|4)$
supergroup manifold leads to the theory of massless higher spin
fields in the $AdS_4$ background.

In order to show this explicitly let us start with the $OSp(1|4)$
invariant superparticle action of \cite{preit}
\begin{equation} \label{7.1}
S= \int d \tau \Lambda_\alpha \Lambda_\beta\, \Omega^{\alpha
\beta},
\end{equation}
where $\Omega^{\alpha \beta}$ has been written in \p{6.16} and
$\Lambda^\alpha$ is a commuting Majorana spinor similar to
$\lambda^\alpha$.

 As it has been proved in the previous section
the $OSp(1|4)$ supergroup  manifold (being the tensorial extension
of the $AdS_4$ superspace) is  $GL(4)$ flat, i.e. its
Grassmann--even Cartan form can be presented in the form \p{ANS}.
This property essentially simplifies the analysis and the
quantization of the action \p{7.1} since after the redefinition of
the twistor variables
$$
\Lambda_\alpha  = {\cal G}_{\alpha}^{-1\,\alpha^\prime}(X,\theta)
\lambda_{\alpha^\prime}
$$
the action takes the `flat' form
\begin{equation}
S= \int d \tau \lambda_\alpha \lambda_\beta \Pi^{\alpha \beta}
\end{equation}
and can be further rewritten in the pure supertwistor form
\p{6}--\p{8}. Therefore the $OSp(1|4)$ model is classically
equivalent to the superparticle in the flat tensorial superspace.
The group theoretical reason for this is that both superspaces are
realized as a coset superspace ${{OSp(1|8)}\over{GL(4)
\times\!\!\!\!\supset SK}}$. As a consequence, when quantizing the
system on $OSp(1|4)$ one can follow the same lines as in the flat
case. Namely, due to the classical equivalence between the flat
and $OSp(1|4)$ superparticle the quantum states of the latter are
again described by the `twistorial' wave functions \p{31}, but
their `Fourier' transform to the coordinate wave functions on the
$AdS_4$ superspace should be performed in a different way. These
wave functions should now obey unfolded equations for higher spin
fields on $AdS_4$. In other words performing quantization and
deriving the equations of motion one should respect the symmetries
of the original classical superparticle model on the tensorial
extension of $AdS_4$ (see \cite{conflat} for the relevant
discussion).

As an illustration consider first several examples corresponding
to the bosonic $AdS_4$ case. Let us take
 the $AdS_4$ metric in a conformally flat form
\begin{equation}
g_{mn} = e^{\rho(x)} \eta_{mn},
\end{equation}
where\footnote{For simplicity we put the radius of the anti de
Sitter space equal to one.} $\rho (x) = \ln{
\frac{4}{(1-x^2)^2}}$, and the corresponding vielbeins and the
spin connection are
$$
e_m^a = e^{\frac{\rho(x)}{2}} \delta^a_m , \quad
 e_a^m = e^{- \frac{\rho(x)}{2}} \delta^m_a,
$$
\begin{equation}
 \omega_{m, ab}= \frac{1}{2} (\eta_{a m}  \partial_b \rho(x)
- \eta_{b m}  \partial_a \rho(x)), \quad \omega_{c, ab} = e^m_c
\omega_{m, ab} = e^{-\frac{\rho(x)}{2}} \omega_{c, ab}.
\end{equation}
For our purposes it is useful to rewrite these expressions in the
two component Weyl spinor notation:
\begin{equation}
 e_{M \dot M}^{A \dot A} = e^{\frac{\rho(x)}{2}} \delta_M^A
\delta_{\dot M}^{\dot A }, \quad \omega_{A \dot A, BC} = -
\frac{1}{4} e^{- \frac{\rho(x)}{2}} (\epsilon _{BA} \partial
_{\dot A C} \rho(x) + \epsilon _{CA} \partial _{\dot A B}
\rho(x)),
\end{equation}
then the $AdS_4$ covariant derivative is defined as
\be
D_{M \dot M} \Phi_{A \dot A}= -\frac{1}{2} \sigma^m_{M \dot M}
\partial_m \Phi_{A \dot A} +\omega_{M \dot M, A}{}^{B}\Phi_{B \dot
A}+ \omega_{M \dot M, \dot A}{}^{\dot B}\Phi_{A \dot B}\,.
\ee

\subsubsection*{s=0}
By the analogy with Section 5 we take the following ansatz for the
scalar field wave function on $AdS_4$
\be\label{79}
\phi(x)={1\over {2\pi}}\int\,d^2\lambda \,d^2\bar\lambda\,
 e^{ix^mP_m -{1\over 2} \rho (x)}\,\phi_0(\l_A\bar\lambda_{\dot A}),
\ee
which satisfies the following field equation
\begin{equation}\label{ms}
(D_m D^m +2) \phi(x) =0\,,
\end{equation}
i.e. the conformally invariant equation for a massless scalar
field propagating on the four dimensional anti de Sitter space
\cite{F1, Me, JB1,conflat}, $D_m=\partial_m+\omega_m$ being the
$AdS_4$ covariant derivative.

The form of the equation \p{ms} is in a full correspondence with
our previous discussion. When considering the quantum version of
the classical constraint \p{5.1} the momentum operator $P_m = -i
\frac{\partial}{\partial x^m}$ becomes non Hermitian in the anti
de Sitter background and one should consider instead the covariant
Laplacian, while because of the quantum ordering ambiguity, the
value of the ``effective mass" appearing in \p{ms} can be fixed by
the requirement of the conformal invariance of the mass shell
equation \cite{F1}.

\subsubsection*{s=1/2}
We take the ansatz for the wave function in the form
\be\label{80}
\psi_{A}(x)={1\over{2\pi}}\,\int\,d^2\lambda
\,d^2\bar\lambda\,e^{ix^mP_m - {3\over 4} \rho(x)}\,\, \lambda_A\,
\bar\lambda_{\dot A}\,\bar\phi^{\dot A}(\l\bar\lambda)\,, \quad
 P_m=2\lambda\sigma_m\bar\l\,.
\ee
which is an $AdS_4$ generalization of the corresponding flat wave
function \p{35}. The value $-{3\over 4}$ of the $\rho(x)$ factor
being now fixed by the requirement for \p{80} to satisfy the
$AdS_4$ Dirac equation
\be
e^{M \dot M}_{A\dot A}\,D_{M \dot M}\,\psi^{ A}(x)=0.
\ee

The procedure described  for the spin $1/2$  field is
straightforwardly generalized to fields of arbitrary (half)integer
spin $s=n/2$. The
 wave function of the form
$$
\psi_{A_1,...A_{2s}}(x)=
$$
\be\label{81}
={1\over{2\pi}}\,\int\,d^2\lambda \,d^2\bar\lambda\,e^{ix^{A\dot
A}\lambda_A\bar\lambda_{\dot A} - \frac{1+ s}{2} \rho (x)}\,\,
\lambda_{A_1}\, ... \lambda_{A_{2s}} \bar\lambda_{\dot A_1}\, ...,
\bar\lambda_{\dot A_{2s}}\, \bar\phi^{\dot A ... \dot
A_{2s}}(\l\bar\lambda),
\ee
satisfies the higher spin field equation
\be
e^{M \dot M}_{A_1\dot A}\,D_{M \dot M}\,\psi^{ A_1 ,...,
A_{2s}}(x)=0.
\ee

We are now in a position to establish the relationship of the
quantized particle on the group manifold $Sp(4)$ with the unfolded
dynamics of higher spin fields in $AdS_4$. In the unfolded
formulation \cite{V,vas,misha} a field with the spin $s$ is
described by an infinite chain of equations
$$
D_{M \dot M} C_{A_1... A_{n+2s};\, \dot A_1... \dot A_n}(x) =
$$
\be \label{ch}
=e_{M \dot M}^{A \dot A} C_{A_1,... A_{n+2s}, A; \dot A_1... \dot
A_n, \dot A}(x) -n (n+2s) e_{M \dot M,\, \{A_1 \dot A_1} C_{A_2...
A_{n+2s}; \dot A_2... \dot A_n \}}(x)
\ee
and by their complex conjugate. The scalar field is described by
the set of fields with an equal number of dotted and undotted
indexes.
 The fields $C_{A_1... A_{2s};\,0}(x)$, $C_{0;\,\dot A_1...
A_{2s}}(x)$ and $C_{0;0}$ (for the scalar field) are physical and
all the other are auxiliary. For spin $s\neq 0$ one obtains from
the first equation in the chain \p{ch} the dynamical equation of
the physical field
\be
e^{M \dot M}_{A_1\dot A}\,D_{M \dot M}\,C^{ A_1 ... A_{2s}}(x)=0
\ee
while the other equations imply no further conditions on the
physical field but just express the auxiliary fields via
derivatives of the physical field. For the scalar field the
situation is slightly different. From the first two equations in
the chain \p{ch} one obtains the Klein--Gordon equation for the
field $C_{0,0}(x)$, while the other equations express the
auxiliary fields in terms of the higher covariant derivatives of
the basic scalar field.

Now one can explicitly see that by identifying the wave function
\p{79} with $C_{0,\,0}(x)$, the wave function \p{81} with
$C_{A_1...A_{2s}}(x)$ and  $D_{\{A \dot A}^n C_{A_1...
A_{2s}\}}(x)$ with the auxiliary fields
  one obtains the complete correspondence between the
superparticle model on $Sp(4)$ and the unfolded massless higher
spin dynamics in the $AdS_4$ space.

 We should note that  higher
spin plane wave solutions on $AdS_4$ have been obtained in
\cite{BV} along with  the corresponding generating functional
$$
C_{\rm plane}(y,\bar y|x) =
$$
\be \label{gf}
=c_0\,exp\left\{i (y_A \bar y_{\dot A}
+ \l_A \bar \l_{\dot A})x^{A \dot A} - \frac{\rho(x)}{2} +
 (1-x^2)^{\frac{1}{2}}(y^{A}\l_{A} + \bar y^{\dot A}\bar \l_{\dot
 A})\right\}, \quad
\ee
$$
C^{\rm plane}_{A_1...A_{2s}}(x)={\partial\over{\partial
y^{A_1}}}\cdots {\partial\over{\partial y^{A_{2s}}}}\,C_{\rm
plane}(y,\bar y|x)|_{y=\bar y = 0}\,,
$$
 where $c_0$ is a constant and the auxiliary oscillator variables $y^A$ and
$\bar y^{\dot A}$ satisfy the Moyal star product commutation
relations \cite{misha}
\be
[y^A\,,y^B] = 2i\varepsilon^{AB}\,, \qquad [\bar y^{\dot A}\,,\bar
y^{\dot B}] = 2i\varepsilon^{\dot A\dot B}\,.
\ee
The generating functional for the generic solution \p{81} of the
higher spin equations \p{ch} on $AdS_4$ is obtained by replacing
$c_0$ in \p{gf} with
$$
\Phi(\lambda_A,\,\bar\l_{\dot A})= \sum_{n=0}^\infty\,\left[
\lambda_{A_1}\cdots\lambda_{A_n}\,\phi^{A_1\cdots
A_n}(\lambda_A\bar\lambda_{\dot A})+ \bar\lambda_{\dot
A_1}\cdots\bar\lambda_{\dot A_n}\bar\phi^{\dot A_1\cdots \dot
A_n}(\l_A\bar\lambda_{\dot A})\right]\,,
$$
and integrating over $\lambda_A$ and $\bar\lambda_{\dot A}$
$$
C(y,\bar y|\,x)=
$$
\be\label{716}
 ={1\over {2\pi}}\int\,d^2\l\,d^2\bar\l\,
 \Phi(\l,\bar\l)\,exp\left\{i (y_A \bar y_{\dot A}
+ \l_A \bar \l_{\dot A})x^{A \dot A} - \frac{\rho(x)}{2} +
 (1-x^2)^{\frac{1}{2}}(y^{A}\l_{A} + \bar y^{\dot A}\bar \l_{\dot
 A})\right\}\,.
\ee

The connection between the unfolded formulation of higher spin
fields in $AdS_4$ and the quantum spectrum of the particle on the
group manifold $Sp\,(4)$ established above can be generalized to
the supersymmetric case. The generating functional \p{716} will
now also depend on the Grassmann--odd supertwistor variable
$\chi$. For example, this can be done in the framework of a
minimal $N=1$ higher spin field theory which corresponds to the
$OSp\,(1|4)$ supersymmetry \cite{ESZ1,ESZ2}. In this formulation
the bosonic part of the generating functional contains  fields
with only even values of spin, while the wave functions of the
superpartners are related to each other by acting on the bosonic
(fermionic) part of the generating functional $C(y,\bar y|x)$ with
the supersymmetry generator $\cal Q$ \p{6.55}
 which in the unfolded formulation is
realized as ${\cal Q}_\alpha = y_\alpha (\chi + \bar \chi)$ (see
Sections 3 and 6). Thus according to \cite{ESZ1, ESZ2} one can
obtain the supersymmetric description of the fields \p{81} grouped
into irreducible $AdS_4$ supermultiplets.

\section{Conclusion}
We have demonstrated that the quantization of the superparticle
propagating in the flat tensorial superspace and on $OSp\,(1|4)$,
which are different cosets
${{OSp(1|8)}\over{GL(4)\times\!\!\!\!\supset SK}}$,  produces the
free massless higher spin field theory in flat and $AdS_4$
superspaces, respectively. An important property of $OSp(1|4)$,
and in general of $OSp(1|2n)$, which we have revealed and which
has allowed us to find the explicit spectrum of quantum higher
spin states of the superparticle on $AdS_4$ is the $GL(2n)$
flatness of these supergroup manifolds. It would be of interest to
analyse whether other groups and supergroups possess this
property. The connection with the unfolded dynamics of higher spin
fields has been achieved for both, the flat and $AdS_4$
superbackground by taking the appropriate ``Fourier''
transformation of the wave function of the quantized
superparticle.

The fact that the quantization of the superparticle dynamics on
the tensorial superspace
${{OSp(1|8)}\over{GL(4)\times\!\!\!\!\supset SK}}$ results in the
dynamics of massless higher spin fields in an associated 4D
superspace--time is in a complete correspondence with the results
of \cite{misha}, where an alternative quantization procedure was
applied, and with \cite{misha1} where it has been shown that the
requirement of the causality and locality of the theory on the
tensorial space singles out the 4D space--time as a subspace of
the tensorial space on which the dynamical higher spin fields are
localized.

The natural development of these results would be to consider the
dynamics of the superparticle in higher dimensional and curved
superbackgrounds and to study the possibility of introducing
interactions of the higher spin fields in this way.

Let us also note that the explicit $GL(2n)$ flat representation of
the Cartan forms of the supergroups $OSp(1|2n)$ found in Section 6
can be useful for many applications where these supergroups
appear, for instance, for the analysis of a conjectured
$OSp(1|32)$ and $OSp(1|64)$ structure of M--theory (see
\cite{conflat} for references).
\\
\\
{\bf Acknowledgements.} The authors are grateful to I. Bandos, E.
Ivanov, P. Pasti, M. Tonin and M. Vasiliev for interest to this
work and useful discussions. This work was partially supported by
the Grants 1010073 and 7010073 from FONDECYT (Chile) (M.P. and
D.S.), by the Grant N 383 of the Ukrainian State Fund for
Fundamental Research (D.S.), by the INTAS Research Project N
2000-254 and by the European Community's Human Potential Programme
under contract HPRN-CT-2000-00131 Quantum Spacetime (D.S. and
M.T.).

\bigskip
{\bf Note added}. When this article was ready for publication, the
paper \cite{dv} appeared on the net in which, in particular,
Cartan forms and solutions of the free massless field equations in
the tensorial AdS (super)space, which generalize \p{gf}, have been
constructed using a star product realization of the $osp\,(N|2n)$
superalgebra. One can assume that the construction of \cite{dv}
should essentially simplify with the use of the $GL(2n)$ flatness
of the $OSp\,(1|2n)$ Cartan forms found in the present paper.

\end{document}